# Calculation of the Electric and Magnetic Root Mean Squared Radiuses of Proton Based on MIT Bag Model

**Maryam Momeni Feili . Mahvash Zandi**
Islamic Azad University, Khorramabad branch
momenifeyli@yahoo.com , mahvash.zandi.n@gmail.com

*Abstract*— **The electric and magnetic bag radiuses of the proton can be determined by MIT bag model based on electric and magnetic form factors of the proton. Also we determined electric and magnetic root mean squared radiuses of the proton, using of bag radius and compared with other results suggests a siutable compatibility.**

## I. INTRODUCTION

While the proton and neutron are thought of as the basic building blocks of visible matter, they are, in fact, complicated bound states of quarks and gluons, held together by the strong interactions described by Quantum Chromodynamics (QCD). The forces binding quarks together are so powerful that it is impossible to simply pluck a single quark from a proton; attempting to do so requires energies so large that new quarks and gluons are "ripped" out of the vacuum, forming new bound states (hadrons) around the quark one is attempting to isolate. Therefore, studying the interactions of QCD involves careful examination of the internal structure of bound states of quarks to isolate information that is directly sensitive to the underlying quark degrees of freedom. Because of this, the proton plays an important dual role as both a basic building block of visible matter and the most accessible bound state of QCD, and as such, has been a primary focus for generations of nuclear physicists [1].

Electric-magnetic form factors provide valuable information about the structure of hadrons and the strong interaction dynamics. At low momenta, they directly probe the electric and magnetic charlge distribution inside the hadron. In general, the form factors are related to the elastic amplitude for a given hadron to absorb a virtual photon. Thus, one can access the interaction responsible for the recombination of the partons in to the hadron.

The electro-magnetic form factors of the nucleon are currently subject to a renewed experimental interest. At low momenta, the proton electric and magnetic form factors can be very well described by the dipole fit:

$$G_D(Q^2) = \frac{1}{\left(1 + \frac{Q^2}{0.71}\right)^2} \tag{1}$$

Where $Q^2$ is the four-momentum transfer squared [2,3].

The positively charged proton is composed of $(u,u,d)$-quarks; it is non-point-like with some non-zero charge distribution and electromagnetic interactions which manifests on electromagnetic structure.

The size of the proton charge and magnetic distributions is described by electric and magnetic root mean squared radiuses in specific (Breit) reference frame [4]:

$$r_{EP} = \sqrt{\langle r_{EP}^2 \rangle} \quad , \quad r_{MP} = \sqrt{\langle r_{MP}^2 \rangle}$$

In this paper we calculate electric and magnetic root mean squared radiuses of proton by MIT bag model.

## II. ELECTRIC AND MAGNETIC BAG RADIUSES

In MIT bag model, It is supposed that the region of space called "bag" including hadron's fields which are fixed [5]. Under the static cavity approximation, the bag surface is spherical and all valence quarks are in the lowest eigen mode. The electric and magnetic form factors for the proton can be written as:

$$G_{EP}(Q^2) = \int_0^{R_E} 4\pi r^2 j_0(Qr)\left[g^2(r)+f^2(r)\right]dr \quad (2)$$

$$G_{MP}(Q^2) = 2m_P \int_0^{R_M} 4\pi r^2 \frac{j_1(Qr)}{Q}\left[2g(r)f(r)\right]dr \quad (3)$$

Where $R_E$ and $R_M$ are the electric and magnetic bag radiuses, and $j_l(x)$ refers to the spherical Bessel function. For a massless quark, the quark wave functions are given as [6]:

$$g(r) = Nj_0(\omega R/r) \quad (4)$$

$$f(r) = Nj_1(\omega R/r) \quad (5)$$

With $\omega = 2.0483$ and $N^2 = \dfrac{\omega}{8\pi R^3 j_0^2(\omega)(\omega-1)}$.

In other words, The distributions of the proton's charge and magnetization are encoded in the elastic electromagnetic form factors, which are measured in elastic electron–proton scattering. The elastic e–p scattering cross section depends on the two elastic form factors, $G_E(Q^2)$ and $G_M(Q^2)$, which depend only on the square of the four-momentum transfer, and which describe the difference between scattering from a structureless particle and an extended, complex object. By measuring the form factors, we probe the spatial distribution of the proton charge and magnetization, providing the most direct connection to the spatial distribution of quarks inside the proton.

In electron scattering, the electron interacts with the proton via exchange of a virtual photon, characterized by its momentum transfer, $Q^2$. At low $Q^2$, the long wavelength photon is sensitive to the size and large scale structure of the proton, while at high $Q^2$, the photon probes the fine details of the structure, as illustrated in Figure 1 (left panel). In the Born (single photon exchange) approximation, the elastic e–p cross section is proportional to the reduced cross section:

$$\sigma_R = \left[(Q^2/4m^2)G_M^2(Q^2) + \varepsilon G_E^2(Q^2)\right] \quad (6)$$

where $m$ is the proton mass, $\theta$ is the electron scattering angle, and $\varepsilon^{-1} = 1+2(1+Q^2/4m^2)\tan^2(\theta/2)$. The presence of the angle-dependent factor $\varepsilon$ multiplying $G_E$ allows a Rosenbluth separation of the form factors by examining the $\varepsilon$ dependence of $\sigma_R$ at fixed $Q^2$ [1].

Now, we calculate the electric and magnetic form factors from the measured cross-section published in Ref. [7] by using Rosenbluth separation as shown in table (I). According to calculated electric and magnetic form factors, the electric and magnetic radiuses of the bag can be obtained based on the eqs. (2), (3) at each value of $Q^2$ [8]. The results are shown in the table (I) and figures (2) and (3). In these calculations, to obtain the static electric and magnetic radius of bag, $R_{E,M}$, the limit value of electric and magnetic bag radiuses can be calculated in $(Q^2 \to 0)$ based on the fits of the graph in figure (2) and (3) which equal to:

$$R_{0E} = 0.9886 \text{fm} \quad , \quad R_{0M} = 1.1563 \text{fm}$$

### III. ELECTRIC ROOT MEAN SQUARED RADIUS

As one of the basic nuclear properties, the root-mean-square (rms) charge radii of nuclei are of great importance for the study of nuclear structures and nucleus-nucleus interaction potentials. On one hand, the rms charge radii of nuclei can be self-consistently calculated by using microscopic nuclear mass models such as the Skyrme Hartree-FockBogoliubov (HFB) model and the relativistic mean-field (RMF) model [9]. On the other hand, the rms charge radii of nuclei are also frequently described by using mass- and isospin-dependent (or charge-dependent) phenomenological formulas. Although these microscopic and phenomenological models can successfully describe the nuclear charge radii of most nuclei, the parabolic charge radii trend in the Ca isotope chain due to the shell closure of $N = 20$ and $N = 28$ cannot be reasonably well reproduced [10]. The electric root mean squared of proton in MIT bag model is defined as [5]:

$$\langle r^2 \rangle_{EP} = \frac{1}{e} \int d^3r_1 \int d^3r_2 \int d^3r_3 \Psi_p^\dagger \left( \hat{Q}_i \hat{r}_i^2 \right) \Psi_p \quad (7)$$

Here $\Psi_p$ denotes the proton wave function and $\hat{Q}_i$ is charge operator. Since the $SU(6)$ wave function of the proton is symmetric under permutations of the indices 1, 2 and 3, we have:

$$\int d^3r_1 \Psi_p^\dagger \left( \hat{Q}_1 \hat{r}_1^2 \right) \Psi_p = \int d^3r_2 \Psi_p^\dagger \left( \hat{Q}_2 \hat{r}_2^2 \right) \Psi_p = \int d^3r_3 \Psi_p^\dagger \left( \hat{Q}_3 \hat{r}_3^2 \right) \Psi_p \quad (8)$$

We insert $\Psi_p$ in Eq. (7), then:

$$\langle r^2 \rangle_{EP} = \frac{1}{6e} \int d^3r_1 \int d^3r_2 \int d^3r_3$$
$$\left[ 2u\uparrow(1)u\uparrow(2)d\downarrow(3) - u\uparrow(1)d\uparrow(2)u\downarrow(3) - d\uparrow(1)u\uparrow(2)u\downarrow(3) \right]^\dagger \quad (9)$$
$$\times \left[ Q_1 r_1^2 + Q_2 r_2^2 + Q_3 r_3^2 \right]$$
$$\left[ 2u\uparrow(1)u\uparrow(2)d\downarrow(3) - u\uparrow(1)d\uparrow(2)u\downarrow(3) - d\uparrow(1)u\uparrow(2)u\downarrow(3) \right]$$

All quarks are assumed to be massless and the same state. We do not distinguish between up and down quarks. Therfore:

$$r_1^2 = r_2^2 = r_3^2$$
$$Q_1 r_1^2 + Q_2 r_2^2 + Q_3 r_3^2 \rightarrow (Q_1 + Q_2 + Q_3) r_1^2 = e r_1^2 \quad (10)$$

Now, the integrals over $r_2$ and $r_3$ can easily be evaluated by orthogonality relation. Then:

$$\langle r^2 \rangle_{EP} = \int d^3 r_1 r_1^2 \frac{1}{6} \left[ 4u\uparrow^\dagger(1)u\uparrow(1) + u\uparrow^\dagger(1)u\uparrow(1) + d\uparrow^\dagger(1)d\uparrow(1) \right] \quad (11)$$

The wave functions of $u$ and $d$ yield the same contributions, and Eq. (11) is simplified to:

$$\langle r^2 \rangle_{EP} = \int d\Omega \int dr_1 r_1^4 u\uparrow^\dagger(1) u\uparrow(1) \quad (12)$$

Next we insert wave function of quarks obtain:

$$\langle r^2 \rangle_{EP} = \frac{\omega}{2R^3(\omega-1)j_0^2(\omega)} \int_0^{R_E} dr r^4 \left[ j_0^2(Er) + j_1^2(Er) \right] =$$
$$\frac{-ER_E \sin(ER_E)\cos(ER_E) + \frac{1}{3}E^3 R_E^3}{2E^2(ER_E - 1)\sin^2(ER_E)} = 0.53 R_E^2 \quad (13)$$

In the last step, we have inserted the numerical value for $\omega = ER_E = 2.0428$. According to the calculated static electric radius, the numerical value of electric root mean squared radius can be obtained and compared with the results obtained by others. As it can be seen in table (II).

## IV. MAGNETIC ROOT MEAN SQUARED RADIUS

An accurate knowledge of the magnetic radius of the proton is fundamental for a deep understanding of its internal structure. We have also performed the above procedure to find the proton magnetic radius. We can define the magnetic root mean squared radius of proton in MIT bag model as:

$$\langle r^2 \rangle_{MP} = \frac{1}{2e\mu_P} \int d^3r_1 \int d^3r_2 \int d^3r_3 \Psi_p^\dagger (\hat{\mu}\hat{r}_i) \Psi_p \qquad (14)$$

$$\hat{\mu} = \frac{\hat{Q}_i}{2} \hat{r}_i \times \alpha_i$$

Here $\mu_P$ is magnetic moment of proton. We insert $\Psi_p$ in Eq. (14). Then:

$$\langle r^2 \rangle_{MP} = \frac{1}{4e\mu_P} \int r_1 d^3r_1 [10 u\uparrow^\dagger(1)\left(\frac{\hat{Q}_1}{2}\hat{r}_1 \times \alpha_1\right) u\uparrow(1) +$$
$$2u\downarrow^\dagger(1)\left(\frac{\hat{Q}_1}{2}\hat{r}_1 \times \alpha_1\right) u\downarrow(1) + 4d\uparrow^\dagger(1)\left(\frac{\hat{Q}_1}{2}\hat{r}_1 \times \alpha_1\right) d\uparrow(1) \qquad (15)$$
$$+2d\uparrow^\dagger(1)\left(\frac{\hat{Q}_1}{2}\hat{r}_1 \times \alpha_1\right) d\uparrow(1)]$$

In the Eq. (15) we use symmetric wave function of the proton under permutations of the indices 1, 2, and 3. Also the integrals over $r_2$ and $r_3$ calculated by orthogonality relation.

Now, we insert the value of quark charges, wave function of quarks and $\alpha_i = \begin{pmatrix} 0 & \sigma_i \\ \sigma_i & 0 \end{pmatrix}$. Where $\sigma_i$ is pauli matrices in Eq. (15). Then:

$$\langle r^2 \rangle_{MP} = \frac{\omega}{6\mu_P R^3 (\omega-1) j_0^2(\omega)} \int_0^{R_M} dr r^4 \left[ j_0(Er) j_1(Er) \right] =$$
$$\frac{R_M^2}{12\mu_P} \left[ \frac{ER_M^2 \cos^2(ER_M) + (2ER_M - 1)\sin^2(ER_M)}{E^2 R_M^2 (ER_M - 1)\sin^2(ER_M)} \right] = 0.39 R_M^2 \qquad (16)$$

In the last step we use $\omega = ER_M = 2.0428$ and $\mu_P = 2.79$.

According to the calculated static magnetic radius, the numerical value of magnetic root mean squared bag radius can be obtained in $(Q^2 \to 0)$ based on the fits the graph in figure (2).

## V. CONCULTION

The electric and magnetic root mean squared radiuses of proton are obtained based on MIT bag model and calculated numerical value of them by using of the static electric and magnetic radius of bag.

The electric and magnetic radiuses of bag calculated by using of electric and magnetic form factors of proton. It is seem that the value of $R_E$ and $R_M$ is in reverse proportion to the increase of $Q^2$, since $Q^2$ increases cause an external pressure imposed on the surface of the bag and make it contract, hence, its radiuses decreases.

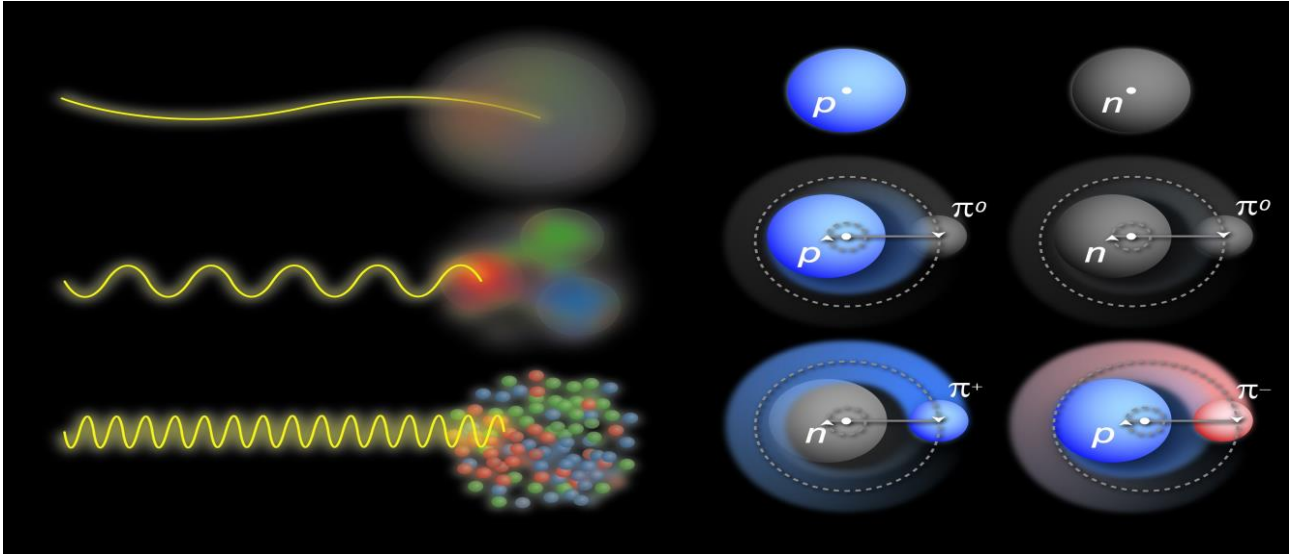

**Fig.1** Left: Illustration of the structure probed in electron–proton scattering at various energy scales. At low energy, the virtual photon probes the proton at large distance scales and is sensitive to the size and large-scale structure. At higher energies, the photons probe shorter distance scales, probing the constituent quarks or the sea of quark-antiquark pairs. Right: Illustration of the impact of the "pion cloud" to the charge distribution of the proton (left) and neutron (right) charge radii. Blue (red) indicates positive (negative) charge, and presence of a virtual pion–nucleon has two main effects. First, the motion of the system yields a 'smearing' of the nucleon charge distribution due to its center-of-mass motion. Second, the charge pion associated with the $n \rightarrow p + \pi^-$ fluctuation yields a negative contribution at large distances. Credit: Joshua Rubin, Argonne National Laboratory[1].

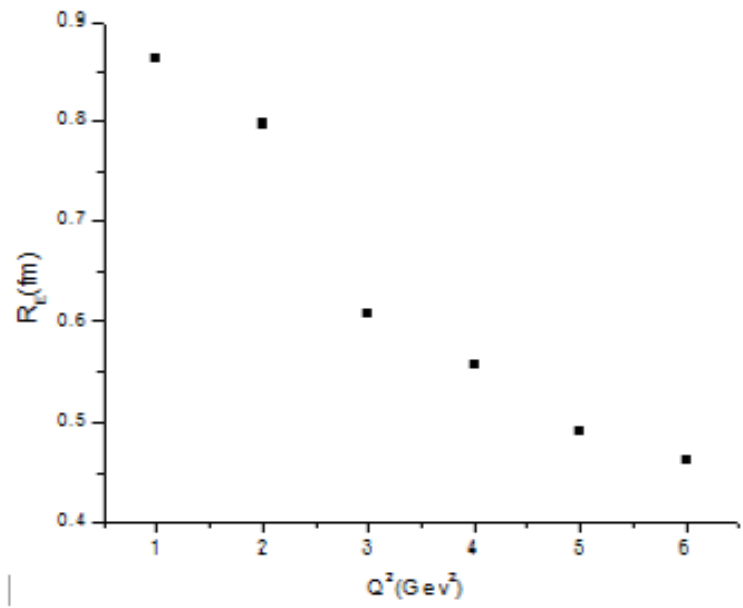

**Fig.2** The electric bag radius according to values of $Q^2$ Ref. [7].

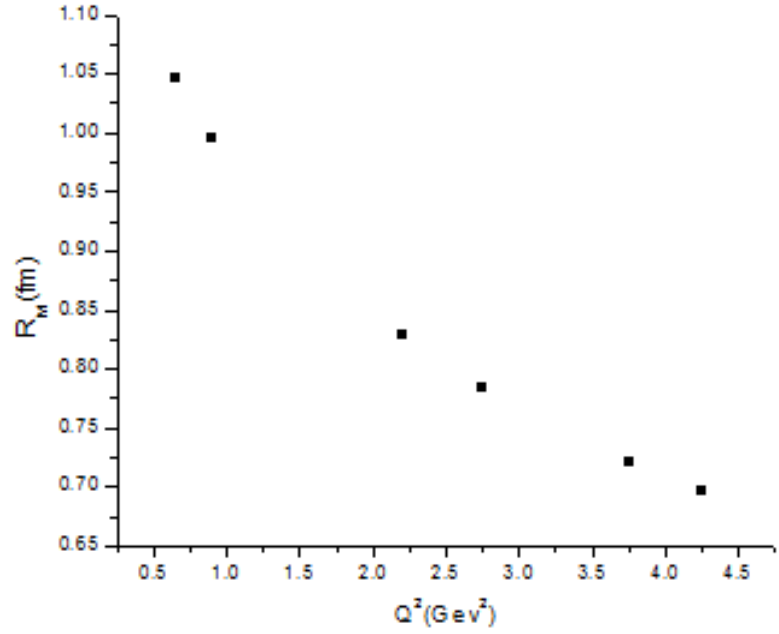

**Fig.3** The magnetic bag radius according to values of $Q^2$ Ref. [7]

**Table I** Calculate Electric and Magnetic Form Factors of Proton and Electric and Magnetic Radius of Bag

| $Q^2 (GeV/c)^2$ | $G_E^P$ | $G_M^P$ | $R_E (fm)$ | $R_M (fm)$ |
|---|---|---|---|---|
| 0.65 | 0.2687 | 0.735 | 0.863 | 1.047 |
| 0.9 | 0.1941 | 0.559 | 0.797 | 0.995 |
| 2.2 | 0.0557 | 0.174 | 0.608 | 0.829 |
| 2.75 | 0.0382 | 0.124 | 0.558 | 0.784 |
| 3.75 | 0.0236 | 0.076 | 0.490 | 0.721 |
| 4.25 | 0.6202 | 0.058 | 0.462 | 0.696 |
| 5.25 | 0.0155 | 0.040 | 0.419 | 0.653 |

**Table II** Our Results and Comparison With Other

|  | Our | Ref. [4] | Ref. [10] | Ref. [11] | Ref.[12] |
|---|---|---|---|---|---|
| $\langle r_E^2 \rangle_P^{1/2} (fm)$ | 0.719 | 0.848 | 0.735 | 0.84 |  |
| $\langle r_M^2 \rangle_P^{1/2} (fm)$ | 0.722 | - | 0.631 | 0.85 | 0.777 |